\documentclass[%
 reprint,
 amsmath,amssymb,
 aps,
]{revtex4-2}

\usepackage{graphicx}% Include figure files
\usepackage{dcolumn}% Align table columns on decimal point
\usepackage{bm}% bold math
\usepackage{color}     % Define colors
\usepackage{amsmath}

\begin{document}

\preprint{APS/123-QED}

\title{An effective and efficient algorithm for the Wigner rotation matrix at high angular momenta}

\author{Bin-Lei Wang}%
\affiliation{School of Physical Science and Technology, Southwest University, Chongqing 400715, China}%

\author{Fan Gao}%
\affiliation{School of Physical Science and Technology, Southwest University, Chongqing 400715, China}%

\author{Long-Jun Wang}
\email{longjun@swu.edu.cn}
\affiliation{School of Physical Science and Technology, Southwest University, Chongqing 400715, China} %Lines break automatically or forced with \\

\author{Yang Sun}
\affiliation{School of Physics and Astronomy, Shanghai Jiao Tong University, Shanghai 200240, China}%

\date{\today}% It is always \today, today,
             %  but any date may be explicitly specified

%%%%%%%%%%%%%%%%%%%%%%%%%%%%%%%%%%%%%%%%%%%%%%%%%%%%%%%%%%%%%%%%%%%%%%%%%%%%%%%%%%%%%%%%%%%%%%%%%%%%%%%%%%%%%%%%%%%
\begin{abstract}
  The Wigner rotation matrix ($d$-function), which appears as a part of the angular-momentum-projection operator, plays a crucial role in modern nuclear-structure models. However, it is a long-standing problem that its numerical evaluation suffers from serious errors and instability, which hinders precise calculations for nuclear high-spin states. Recently, Tajima [Phys. Rev. C 91, 014320 (2015)] has made a significant step toward solving the problem by suggesting the high-precision Fourier method, which however relies on formula-manipulation softwares. In this paper we propose an effective and efficient algorithm for the Wigner $d$ function based on the Jacobi polynomials. We compare our method with the conventional Wigner method and the Tajima Fourier method through some testing calculations, and demonstrate that our algorithm can always give stable results with similar high-precision as the Fourier method, and in some cases (for special sets of $j, m, k$ and $\theta$) ours are even more accurate. Moreover, our method is self-contained and less memory consuming. A related testing code and subroutines are provided  as Supplemental Material in the present paper. 
\end{abstract}

%\keywords{Suggested keywords} %Use showkeys class option if keyword
                               %display desired
\maketitle

%\tableofcontents

%%%%%%%%%%%%%%%%%%%%%%%%%%%%%%%%%%%%%%%%%%%%%%%%%%%%%%%%%%%%%%%%%%%%%%%%%%%%%%%%%%%%%%%%%%%%%%%%%%%%%%%%%%%%%%%%%%%
\section{\label{sec:intro}Introduction}

The microscopic description of collective rotational motion involves  quantum-mechanical treatment of angular momentum, in which the three angular-momentum operators, $\hat j_i$  ($i=x,y,z$), are generators of the Lie algebra of SU(2) and SO(3). The Wigner $D$-matrix, a unitary matrix in an irreducible representation of the groups SU(2) and SO(3), enters into the discussion when functions of angular momentum are transformed by the rotation operator $\hat R(\phi, \theta, \psi) = e^{-i\phi \hat j_z} e^{-i\theta \hat j_y} e^{-i\psi \hat j_z}$ with the Euler angles $(\phi, \theta, \psi)$ \cite{Sun_2022_book}. If the eigenstates of the angular momentum operators are expressed in the spherical basis and labeled by the quantum number $j$ (with $j=0, \frac{1}{2}, 1, \frac{3}{2}, \cdots$) and the projection on the $z$ axis with $2j+1$ quantum numbers labeled as $m$ or $k = -j, -j+1, \cdots, j-1, j$, the Wigner $D$-matrix can be written as, 
\begin{eqnarray} \label{big_D}
  D^j_{mk}(\phi, \theta, \psi) &=& \langle jm| \ e^{-i\phi \hat j_z} e^{-i\theta \hat j_y} e^{-i\psi \hat j_z} \ |jk \rangle \nonumber \\
                               &=& e^{-i(m\phi + k\psi)} d^j_{mk}(\theta),
\end{eqnarray}
where
\begin{eqnarray} \label{small_d}
  d^j_{mk}(\theta) = \langle jm | \ e^{-i\theta \hat j_y} \ | jk \rangle.
\end{eqnarray}
is the key part in the expression, referred to as  Wigner (small) $d$-matrix. As Eqs. (\ref{big_D}) and (\ref{small_d}) are functions of the Euler angles for different sets of quantum numbers $\{j, m, k\}$, they are usually called Wigner $D$- ($d$)-functions, respectively \cite{Angular_book, Tajima_d_PRC_2015}. 

\begin{figure}[htbp]
\begin{center}
  \includegraphics[width=0.49\textwidth]{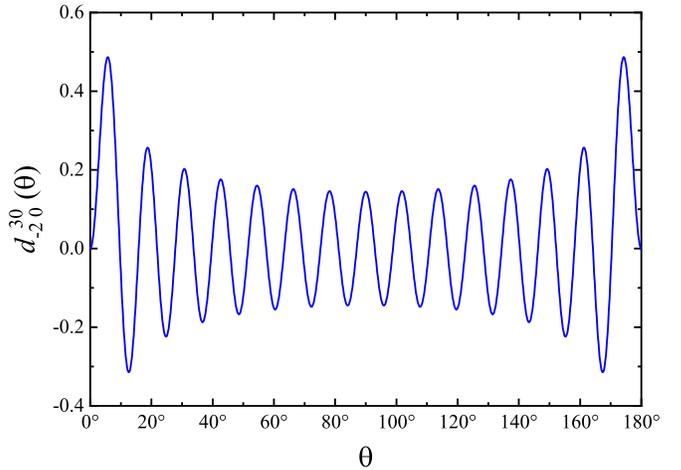}
  \caption{\label{fig:fig_1} (Color online) The highly oscillatory behavior of the Wigner $d^j_{mk} (\theta)$ function with $j=30, m=-2, k=0$.} 
\end{center}
\end{figure}

As its characteristic feature, $d^j_{mk}(\theta)$ is known as an oscillation function of $\theta$. In general, the oscillation frequency of the $d$-function increases rapidly with the angular momentum $j$.  Figure \ref{fig:fig_1} shows the example for $j=30 \hbar$.

%------------------------- Figure:

The Wigner $D$-function plays crucial roles in many discussions of modern physics. As a set of orthogonal functions of the Euler angles, the $D$-function can be used to expand other functions. The $D$-function is related to some other well-known functions, as for instance, $ D^j_{m0}(\phi, \theta, \psi) = \sqrt{\frac{4\pi}{2j+1}} Y^\ast_{jm}(\theta, \phi)$ and $D^j_{00}(\phi, \theta, \psi) = P_j (\cos \theta)$, where $Y_{jm}$ ($P_j$) is the spherical harmonic function (Legendre polynomial) \cite{Angular_book, Suhonen_book}. Furthermore, irreducible tensors can be well defined with the help of the Wigner $D$-function. Therefore, the Wigner $D$-function is indispensable in the study of modern physics, for example in nuclear physics \cite{Ring_many_body_book}, quantum metrology \cite{tmp_PRL_2007, tmp_PRL_2008} and many other fields \cite{tmp_PRA_2007, tmp_EPJD_2012}. Especially in theoretical calculations with the nuclear beyond-mean-field methods \cite{PSM_review, PSM-Sun, PSM_Sun2, Niksic_PRC_2009, Niksic_2011_PPNP, Egido_2016, Yao_PPNP}, where angular-momentum projection serves as an important ingredient, the Wigner $D$-function is the cerntral part of the angular-momentum projector \cite{PSM_review, TPSM_Sheikh_2011_PRC, Zhao_P_W_projection_2016, Calvin_Johnson_2017_PRC, LJWang_2014_PRC_Rapid, LJWang_2016_PRC, LJWang_PLB_2020_chaos, Sheikh_2021_PRC, Babra_2021_PRC, Sheik_2021_EPJA, Rani_2021_EPJA, ZRChen_PRC}. Angular-momentum projected wave functions obtained in modern nuclear theories are applied to the study of $\beta$-decay \cite{ZCGao_2006_PRC_Gamow_Teller, LJWang_2018_PRC_GT}, neutrinoless double-$\beta$ decay \cite{Tomas_2010_PRL_0vbb, Yao_2015_PRC_0vbb, LJWang_current_2018_Rapid}, astrophysical weak process \cite{LJWang_PLB_2020_ec, LJWang_2021_PRL, 93Nb_PRC}, nuclear fission \cite{Fission_Bertsch_Robledo_2019_PRC, Robledo_2018_JPG_review, TD_GCM_2016_cpc, TD_GCM_2018_cpc, TD_GCM_2019_PRC}, and many others.

All the above applications require a numerically accurate and computationally stable evaluation of the small $d$-function. However, due to the presence of many factorials of large numbers in the formula \cite{Angular_book, Feng_d_PRE_2015}, numerical calculations of the $d$-function by the conventional Wigner method suffers from a serious loss of precision at medium and high spins. Although a few remedies have been proposed \cite{Choi_1999, Dachsel_2006, Prezeau_2010}, they still encounters severe numerical instability and/or loss of precision from unclear sources. A few years ago, Tajima \cite{Tajima_d_PRC_2015} proposed a new method for the Wigner $d$-function evaluation based on the Fourier-series expansion. In this method, the precision of the $d$-function is determined by accurate evaluation of the Fourier coefficients. This method turns out to be of a significant improvement for the numerical stability and precision in the $d$-function evaluation. However, those Fourier coefficients still involve many factorials of large numbers so that they have to be evaluated with the assistance of a formula-manipulation software. Besides, those Fourier coefficients take up a lot of memory in the numerical procedure. 

In this work, we propose an alternative method based on the Jacobi polynomials with a stable and high-precision algorithm for the Wigner $d$-function evaluation. We show that our method can achieve a very similar precision and stable result as the Tajima Fourier method, but ours may be more efficient and user-friendly. In Sec. \ref{sec:theory} we provide, step by step, the analytic expressions of the Wigner, Fourier, and Jacobi methods for the Wigner $d$ function. The precision of the Jacobi method is analyzed in details and compared with both the Wigner and Fourier methods in Sec. \ref{sec:resul} and Sec. \ref{sec:integral}. We finally summarize our work in Sec. \ref{sec:sum}.

%%%%%%%%%%%%%%%%%%%%%%%%%%%%%%%%%%%%%%%%%%%%%%%%%%%%%%%%%%%%%%%%%%%%%%%%%%%%%%%%%%%%%%%%%%%%%%%%%%%%%%%%%%%%%%%%%%%
\section{\label{sec:theory}Different methods for the Wigner $d$-function evaluation}

The conventional method for the $d$-function is based on the following Wigner formula  \cite{Angular_book}, i.e. 
\begin{eqnarray} \label{wigner_d}
  d^j_{mk}(\theta) = \sum_{n=n_{\text{min}}}^{n_{\text{max}}} (-1)^n W^{jmk}_n (\theta),
\end{eqnarray}
where 
\begin{subequations}
\begin{eqnarray}
  n_{\text{min}} &=& \text{max}(0, k-m), \\
  n_{\text{max}} &=& \text{min}(j-m, j+k),
\end{eqnarray}
\end{subequations}
and 
\begin{eqnarray}
  W^{jmk}_n (\theta) &=& w^{jmk}_n \left( \cos \frac{\theta}{2} \right)^{2j+k-m-2n} \left( -\sin \frac{\theta}{2} \right)^{m-k+2n}, 
\end{eqnarray}
with
\begin{eqnarray} \label{wigner_w}
  w^{jmk}_n &=& \frac{\sqrt{ (j+m)!(j-m)! (j+k)!(j-k)! }}{(j-m-n)! (j+k-n)! (n+m-k)! n!} .
\end{eqnarray}
It can be seen that the Wigner formula (\ref{wigner_d}) involves a summation over many terms, $W^{jmk}_n$, with alternating signs. Each of these terms includes many factorials of large numbers, especially at medium and high spins, as they grow exponentially with $j$ ($W^{jmk}_n \approx 2^j$). Although cancellation among these terms should finally lead the summation to a normal value for the $d$-function, the procedure would however cause an accumulation of numerical errors. Thus the numerical results from the Wigner formula unavoidably suffers from a serious loss of precision at medium and high spins, except for the neighborhood of $\theta=0$ and $\pi$ \cite{Tajima_d_PRC_2015}. 

The problem is the repeated production and cancellation of large numbers. To avoid the problem, Tajima \cite{Tajima_d_PRC_2015} proposed a new method for the Wigner $d$-function based on Fourier-series expansion, in which the $d$-function can be expressed as
\begin{eqnarray} \label{Tajima_d}
  d^j_{mk}(\theta) = \sum_{\rho = \rho_{\text{min}}}^{j} t^{jmk}_\rho f(\rho \theta).
\end{eqnarray}
In the above formula, $\rho_{ \text{min} }$ could be $0, \ 1/2$ or $1$ depending on the values of $j, m$ and $k$ (see Table I of Ref. \cite{Tajima_d_PRC_2015} for details) and $f$ is $\sin$ ($\cos$) function for $m-k$ being odd (even). In Eq. (\ref{Tajima_d}) the Fourier coefficient reads
\begin{eqnarray}
  t^{jmk}_\rho &=& \frac{2(-1)^{m-k}}{1+\delta_{\rho 0}} \sum_{n=n_{\text{min}}}^{n_{\text{max}}} (-1)^n  w^{jmk}_n \sum_{r=0}^{[\rho-{1 \over 2}p]} (-1)^r {2\rho \choose 2r+p} \nonumber \\
  & & \ \times {1 \over 2\pi} I_{2(j+\rho-n-r)-m+k-p, \ 2(n+r)+m-k+p},
\end{eqnarray}
where $p\equiv |m-k|$ (mod 2), the square brackets are the floor function \cite{Tajima_d_PRC_2015}, ${\rho \choose r} = \rho! / [r! (\rho - r)!]$, and 
\begin{eqnarray}
  I_{\lambda \alpha} = \int_0^{2\pi} \cos^\lambda x \  \sin^\alpha x dx.
\end{eqnarray} 

The Fourier method avoids cancellation among terms with large numbers since each Fourier coefficient, $t^{jmk}_\rho$, is less than or equal to 1. It has indeed much improved the calculation of the $d$-function. However, two factors may limit its application. On one hand, the Fourier coefficient, $t^{jmk}_\rho$, still includes many factorials of large numbers, i.e., $w^{jmk}_n$, so that it has to be calculated by means of formula-manipulation software such as $\mathtt{MAXIMA}$ or \emph{Mathematica} \cite{Tajima_d_PRC_2015}. On the other hand, in practical applications, one has to read  $t^{jmk}_\rho$ from files and store into a memory, which consumes about 70 MB (1.2 GB) space for the case of $j \leqslant 50$ ($j \leqslant 100$) \cite{Tajima_d_PRC_2015}. Recently, Feng \emph{et al.} put forward an exact-diagonalization method to calculate the Fourier coefficients, in which the corresponding precision of the $d$-function decreases a little and the requested space doubles to be about 2.4 GB for the case of $j \leqslant 100$ \cite{Feng_d_PRE_2015}. 

It is thus desired to have an efficient algorithm for the $d$-function evaluation, which, while keeps the high precision as of the Tajima Fourier method, is self-contained, and therefore, user-friendly. Towards this goal, we note that the Wigner $d$-function can be expressed in terms of the Jacobi polynomials  \cite{Angular_book}
\begin{eqnarray} \label{Jacobi_d}
  d^j_{mk}(\theta) &=& \xi_{mk} \left[ \frac{s! (s+\mu+\nu)!}{(s+\mu)! (s+\nu)!} \right]^{1/2} \nonumber \\
          & & \times \left( \sin \frac{\theta}{2} \right)^\mu \left( \cos \frac{\theta}{2} \right)^\nu P^{(\mu, \nu)}_s (\cos \theta) ,
\end{eqnarray}
where $\mu=|m-k|$, $\nu=|m+k|$, $s=j-\frac{1}{2}(\mu+\nu)$ and
\begin{eqnarray}
  \xi_{mk} = \left\{ \begin{array}{ll}
              1 & \text{ if } k \ge m, \\
              (-1)^{k-m} & \text{ if } k < m.
               \end{array} \right. 
\end{eqnarray}
The Jacobi polynomial in Eq. (\ref{Jacobi_d}) can be calculated by its explicit expression  \cite{handbook_of_math_1964}
\begin{eqnarray} \label{Jacobi_explicit}  %---Jacobi_explicit
  P^{(\mu, \nu)}_s (z) = \frac{1}{2^s} \sum_{n = 0}^s {s+\mu \choose n} {s+\nu \choose s-n} (z-1)^{s-n} (z+1)^n , \nonumber \\
\end{eqnarray}
or by the corresponding recurrence relations  \cite{handbook_of_math_1964}
\begin{eqnarray} \label{Jacobi_recurrence}  %---Jacobi_recurrence
  & & 2s (s + \mu + \nu) (2s + \mu + \nu - 2) P^{(\mu, \nu)}_s (z) \nonumber \\
  &=& (2s + \mu + \nu - 1) \Big[ (2s + \mu + \nu) (2s + \mu + \nu - 2) z \nonumber \\
  & & \qquad \qquad \qquad \qquad \qquad \qquad + \mu^2 - \nu^2 \Big] P^{(\mu, \nu)}_{s-1} (z) \nonumber \\
  & & - 2(s + \mu - 1)(s + \nu - 1)(2s + \mu + \nu) P^{(\mu, \nu)}_{s-2} (z) , \nonumber \\
\end{eqnarray}
with 
\begin{subequations} \label{Jacobi_recurrence_2}
\begin{eqnarray} 
  P^{(\mu, \nu)}_0 (z) &=& 1 , \\
  P^{(\mu, \nu)}_1 (z) &=& (\mu + 1) + (\mu + \nu +2) \frac{z-1}{2} . 
\end{eqnarray}
\end{subequations}
It is important to realize that unlike the Wigner method and the Fourier method, the expression in Eq. (\ref{Jacobi_d}) that leads to the Wigner $d$-function does not involve a summation over many terms with large numbers.

%%%%%%%%%%%%%%%%%%%%%%%%%%%%%%%%%%%%%%%%%%%%%%%%%%%%%%%%%%%%%%%%%%%%%%%%%%%%%%%%%%%%%%%%%%%%%%%%%%%%%%%%%%%%%%%%%%%
\section{\label{sec:resul}Error analysis of the Jacobi method}

%------------------------- Figure:
\begin{figure*}[htbp]
\begin{center}
  \includegraphics[width=0.80\textwidth]{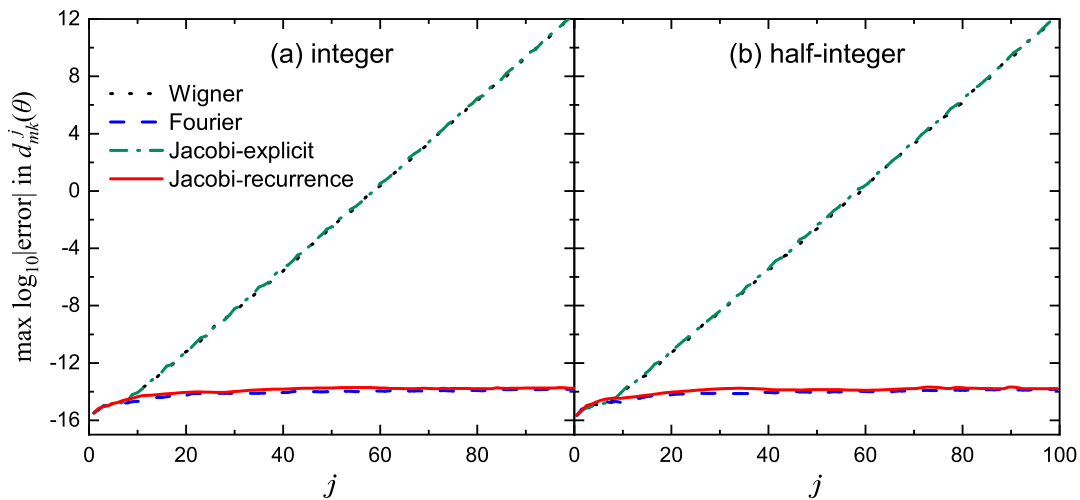}
  \caption{\label{fig:fig_2} (Color online) The common logarithm of the maximum error in the numerical value of the $d$ function $d^j_{mk}(\theta)$ as a function of $j$ for both integer and half-integer cases of $j$, with the Wigner, Fourier and Jacobi methods. The exact values of $d^j_{mk}(\theta)$ are obtained by \emph{Mathematica 12.1} and the maximum error is taken over all the possible values of $m, k$ and $\theta$ for each $j$. See the text for details. } 
\end{center}
\end{figure*}

%------------------------- Figure:
\begin{figure*}[htbp]
\begin{center}
  \includegraphics[width=0.99\textwidth]{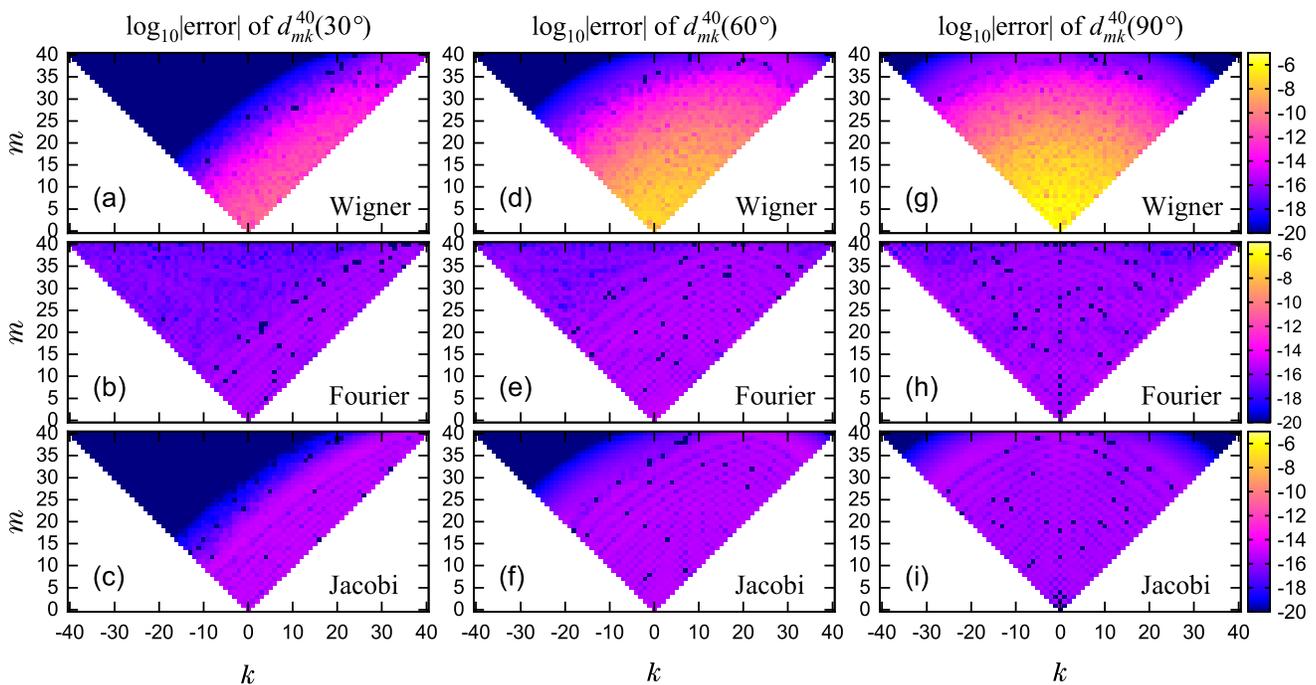}
  \caption{\label{fig:fig_3} (Color online) The common logarithm of the error in the numerical value of the $d$ function $d^j_{mk}(\theta)$ for different values of $m$ and $k$, with $j=40$, $\theta = 30^\circ$, $60^\circ$ and $90^\circ$. The results of the Jacobi algorithm are compared with those of the Wigner and Fourier methods. See the text for details. } 
\end{center}
\end{figure*}

\begin{figure*}[htbp]
\begin{center}
  \includegraphics[width=0.90\textwidth]{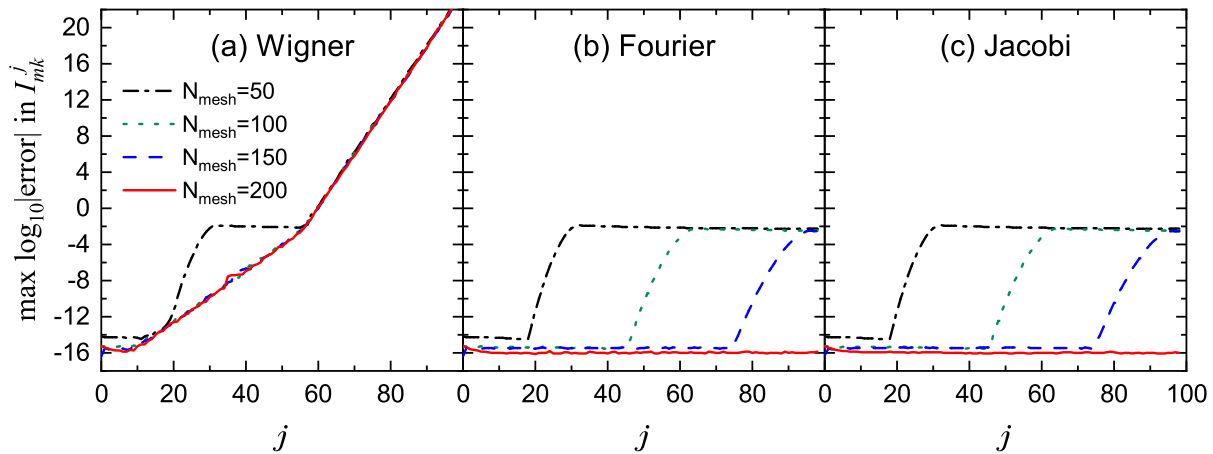}
  \caption{\label{fig:fig_4} (Color online) The common logarithm of the maximum error in the numerical value of the integral $I^j_{mk}$ in Eq. (\ref{eq_integral}) by the Gauss-Legendre quadrature formula as a function of $j$. Cases with different number of mesh points $N_{\text{mesh}}$ by the Jacobi algorithm are compared with those of the Wigner and Fourier methods. The maximum error is taken over all the possible values of $m$ and $k$ for each $j$. See the text for details. } 
\end{center}
\end{figure*}

In this and the next sections we carry out error analysis and discuss precision of the Jacobi method in Eq. (\ref{Jacobi_d}) by comparing with the conventional Wigner method and the recent Fourier method. In Figs. \ref{fig:fig_2} and \ref{fig:fig_3} the absolute errors for the $d$-function from the three different methods are presented, and in Fig. \ref{fig:fig_4} the errors in an integral calculation involving the $d$-function are illustrated. All these results are obtained by a $\texttt{FORTRAN90}$ testing code with standard subroutines for the Wigner, Fourier and Jacobi methods as provided in the Supplemental Material \cite{Suppl_Material}, where in all cases floating-point numbers are adopted as double-precision (64-bit) real number. 

First in Fig. \ref{fig:fig_2}, we show maximum errors of the $d^j_{mk}(\theta)$ calculation as a function of $j$, obtained from the Wigner, Fourier, and Jacobi methods. Results for integer and half-integer $j$'s are illustrated separately. Each error of $d^j_{mk}(\theta)$ is evaluated with respect to the exact value calculated from the formula-manipulation software \emph{Mathematica 12.1} where higher than $10^{-25}$ precision is kept. The maximum of the errors is recorded by considering all $\theta$'s from $0^\circ$ to $180^\circ$ with an increment of $5^\circ$, and for all possible values of $m$ and $k$ with $0 \leqslant m \leqslant j$ and $k \leqslant |m|$ due to the following symmetries, 
\begin{subequations} \label{d_symmetry}
\begin{eqnarray}
  d^j_{mk}(\theta)  &=& (-1)^{m-k} d^j_{-m-k}(\theta), \\
  d^j_{mk}(\theta)  &=& (-1)^{m-k} d^j_{km}(\theta), \\
  d^j_{mk}(\theta)  &=&            d^j_{-k-m}(\theta), \\
  d^j_{mk}(-\theta) &=& (-1)^{m-k} d^j_{mk}(\theta), \\
  d^j_{mk}(-\theta) &=&            d^j_{km}(\theta), \\
  d^j_{mk}(\pi - \theta) &=& (-1)^{j+m} d^j_{m-k}(\theta).
\end{eqnarray}
\end{subequations}

It can be seen from Fig. \ref{fig:fig_2} that the maximum error from the Wigner method increases exponentially with $j$, in the way similar as that of $W^{jmk}_n$ in Eq. (\ref{wigner_d}) of the Wigner formula. Already when $j \geqslant 25$ the error would exceed $10^{-10}$, which may lead to serious numerical problems in applications such as  high-spin calculations in nuclear physics. The origin for loss of precision in the Wigner method is clear. It is caused by the summation over many $W^{jmk}_n$ terms in Eq. (\ref{wigner_d}), where numerical errors are accumulated following a power law of $j$. 

On the contrary, the maximum error from the Fourier method keeps almost constant in a stable way towards high spins, with the precision as high as about $10^{-14}$ even when $j \approx 100$. Although the Fourier method in Eq. (\ref{Tajima_d}) also involves a summation over many terms, each term has very high precision since the corresponding Fourier coefficient $t^{jmk}_\rho$ is calculated by means of the formula-manipulation software $\mathtt{MAXIMA}$ with very high precision and is stored into a memory \cite{Tajima_d_PRC_2015}, so that the accumulation of numerical errors can be avoided. 

For the Jacobi method, as seen from Eq. (\ref{Jacobi_d}), there is no summation over many terms with large numbers and a high-precision evaluation of the $d$-function is then expected. As seen from Fig. \ref{fig:fig_2}, when the Jacobi polynomial is calculated directly by its expression of Eq. (\ref{Jacobi_explicit}), a very similar loss of precision is found for the Jacobi method as the Wigner formula. This is due to the fact that Eq. (\ref{Jacobi_explicit}) involves summation over terms that include factorials of large numbers, which leads to accumulation of numerical errors. However, when the recurrence relations of the Jacobi polynomial in Eqs. (\ref{Jacobi_recurrence}, \ref{Jacobi_recurrence_2}) are adopted, the Jacobi method provides a similar high-precision and stable behavior of the $d$ function as the Fourier method. %The precision turns out to be as high as about $10^{-14}$ at $j \leqslant 100$ in a stable way. 
This clearly suggests that it is the recurrence relations in Eqs. (\ref{Jacobi_recurrence}, \ref{Jacobi_recurrence_2}) that avoid accumulation of numerical errors. Using the recurrence relations to improve numerical precision may be helpful for many other numerical problems. Hereafter, the Jacobi method with the recurrence relations in Eqs. (\ref{Jacobi_recurrence}, \ref{Jacobi_recurrence_2}) is referred to as the Jacobi algorithm for the $d$-function evaluation.

%------------------------- Figure:

To further carry out precision analysis in details, we take the $j=40$ case as an example and show in Fig. \ref{fig:fig_3}  errors of $d^{40}_{mk}(\theta)$ for different values of $m$, $k$ and $\theta$, with $\theta = 30^\circ$, $60^\circ$, $90^\circ$ and $0 \leqslant m \leqslant j$, $k \leqslant |m|$ due to the symmetries in Eq. (\ref{d_symmetry}). The results of the Jacobi algorithm (with the recurrence relations) are compared with those of the Wigner and Fourier methods. It is seen that the Fourier method gives uniformly a $10^{-14} \sim 10^{-15}$ precision nearly irrespective of $m, k$ and $\theta$. The Wigner method, however, leads to a rather unstable precision, depending sensitively on $m, k$ and $\theta$. The precision from the Wigner method could have errors as large as $\sim 10^{-5}$ when $m \sim k \sim 0$ and $\theta = 90^\circ$ as seen from Fig. \ref{fig:fig_3}(g), or it could be very accurate, with the precision as high as $10^{-20}$ when $m \sim j$, $k \sim -j$, $\theta=30^\circ$ and $60^\circ$ (see Fig. \ref{fig:fig_3}(a) and (d)) or $m \sim j$, $|k| \sim j$, $\theta=90^\circ$ (see Fig. \ref{fig:fig_3}(g)), which, for these special cases, is much better than the Fourier method. 

Therefore, In Ref. \cite{Tajima_d_PRC_2015}, Tajima suggested that if a very high precision is needed for the $d$-function evaluation, one should develop a program to switch between the Wigner and Fourier methods with special values of $j, m, k$ and $\theta$. It is now very interesting to compare the results of the Jacobi algorithm (with the recurrence relations) in Fig. \ref{fig:fig_3}. For each set of $j, m, k$ and $\theta$, the Jacobi algorithm always reproduces the one with the higher precision between the Wigner and Fourier methods. This pleasant feature in the Jacobi algorithm makes it a natural choice for a switcher as discussed in Ref. \cite{Tajima_d_PRC_2015}.

%%%%%%%%%%%%%%%%%%%%%%%%%%%%%%%%%%%%%%%%%%%%%%%%%%%%%%%%%%%%%%%%%%%%%%%%%%%%%%%%%%%%%%%%%%%%%%%%%%%%%%%%%%%%%%%%%%%
\section{\label{sec:integral}Error analysis when the Wigner $d$-function is integrated}

According to the Peter–Weyl theorem, the Wigner $D$-functions, $D^j_{mk}(\phi, \theta, \psi)$, form a complete set of orthogonal functions of the Euler angles, and are often used to expand functions that are related to rotation. As the Euler angles are continuous variables the expansion takes the form of integrals with $D^j_{mk}(\phi, \theta, \psi)$ being part of the integrand. Because of the highly oscillatory behavior of the $d$-function, as shown in Fig. \ref{fig:fig_1}, especially at high $j$'s, a high precision in numerical calculations for integrals involving the $d$-function becomes an issue. 

For discussions, let us take an example from the calculation with angular-momentum projection for the symmetry-violated nuclear wave-functions from mean-field calculations. Assuming axial symmetry for the intrinsic states, $|\Phi_k \rangle$, the $d$-function enters into the calculation through the angular-momentum projector,
\begin{eqnarray} \label{projector}
  \hat P_{mk}^j = (j+{1\over 2}) \int_0^\pi d^j_{mk}(\theta) \hat R(\theta) \sin{\theta} d\theta,
\end{eqnarray}
where $\hat R(\theta) = e^{-i\theta \hat j_y}$ is the rotation operator around the $y$-axis. It can be generally shown \cite{PSM_review} that the calculated Hamiltonian matrix elements, for example, takes the form
\begin{eqnarray} \label{PSMelement}
  \int_0^{\pi} d^j_{kk^\prime}(\theta)  \langle \Phi_k | \ \hat H \hat R(\theta) \ | \Phi_{k^\prime} \rangle \sin{\theta} d\theta,
\end{eqnarray}
which is an integral over the Euler angle $\theta$ with essentially two functions in the integrand, $d^j_{kk^\prime}(\theta)$ and the rotated matrix elements $\langle \Phi_k | \ \hat H \hat R(\theta) \ | \Phi_{k^\prime} \rangle$ which is expected to be a smooth function of $\theta$. Due to the highly oscillatory behavior of the $d$-function, its precision may be more important for integrals involving the $d$-function as in many potential applications in nuclear physics, quantum metrology and many other fields in the future.

 In Fig. \ref{fig:fig_4} we take one such integral for discussions and show the absolute value (error) of the following integral
\begin{eqnarray} \label{eq_integral}
  I^j_{mk} = \int_0^\pi d\theta \sin\theta \ d^j_{mk}(\theta) \ d^{j+1}_{mk}(\theta) = 0.
\end{eqnarray}
calculated by the standard Gauss-Legendre quadrature formula with different number of mesh points $N_{\text{mesh}}$. The results of the Jacobi algorithm are compared with those of the Wigner and Fourier methods. It is seen that the error from the Wigner method increases rapidly with $j$ and exceeds $10^{-8}$ at $j \gtrsim 35$, irrespective of $N_{\text{mesh}}$, indicating that the error comes mainly from the $d$-function. By comparison, the error from the Jacobi algorithm and Fourier method depends on $N_{\text{mesh}}$ and the precision of the integral (\ref{eq_integral}) could be as high as $10^{-16}$ for $j \leqslant 100$ if $N_{\text{mesh}}=200$ is taken. This suggests that the Jacobi algorithm for the $d$-function applied in integration calculations can achieve a similar high precision as the Fourier method. The remaining errors in Fig. \ref{fig:fig_4} should then come mainly from the quadrature formula. 

Of course, in numerical calculations and practical applications much more complicated integrands generally appear in integrals, for which a large $N_{\text{mesh}}$ is expected, and causes heavier computational burden. Nevertheless, the results in Fig. \ref{fig:fig_4}(b) and (c) suggest that one has a choice to use smaller numbers of mesh points if states of only lower angular momenta are studied. 

%From Tajima paper: the absolute value of the $d$ function is not greater than one because it is a matrix element of a unitary operator between normalized states.

%%%%%%%%%%%%%%%%%%%%%%%%%%%%%%%%%%%%%%%%%%%%%%%%%%%%%%%%%%%%%%%%%%%%%%%%%%%%%%%%%%%%%%%%%%%%%%%%%%%%%%%%%%%%%%%%%%%
\section{\label{sec:sum}summary}

The Wigner $D$- ($d$)-functions serve as indispensable ingredients for many nuclear-structure models and are important for nuclear physics, quantum metrology and many other fields. Numerical evaluation of the Wigner $d$-function, $d^j_{mk}(\theta)$, from the conventional Wigner method suffers from serious errors and instability, especially at medium and high spins. In this paper we present a high-precision and stable algorithm for evaluation of the Wigner $d$-function. The algorithm is based on the Jacobi polynomial and its recurrence relations.

Compared with the conventional Wigner method, the loss of precision at medium and high spins is avoided in our Jacobi algorithm, with a very high precision $10^{-14} \sim 10^{-15}$ when $j \leqslant 100$. Compared with the recent Fourier method, our Jacobi algorithm avoids the dependence on formula-manipulation softwares and does not need a large memory. With the help from the recurrence relations of the Jacobi polynomial, the Jacobi algorithm always gives the best precision so far irrespective of the values of $j, m, k$ and $\theta$. %which turns out to be effective for numerical evaluation of integrals involving the highly oscillatory $d$ function as well. 
Furthermore, it is self-contained, and therefore, user-friendly.

The Jacobi algorithm could be the most effective algorithm for the Wigner $d$-function evaluation in nuclear physics, quantum metrology, and many other fields in the future. The related $\texttt{FORTRAN90}$ testing code and subroutines for the Wigner and Fourier methods as well as the Jacobi algorithm are provided as a Supplemental Material of the present article.

%%%%%%%%%%%%%%%%%%%%%%%%%%%%%%%%%%%%%%%%%%%%%%%%%%%%%%%%%%%%%%%%%%%%%%%%%%%%%%%%%%%%%%%%%%%%%%%%%%%%%%%%%%%%%%%%%%%
\begin{acknowledgments}
  This work is supported by the Fundamental Research Funds for the Central Universities (Grant No. SWU-KT22050), by the National Natural Science Foundation of China (Grant No. 11905175 and No. U1932206), by the Natural Science Foundation of Chongqing and partially supported by the Key Laboratory of Nuclear Data (China Institute of Atomic Energy).
\end{acknowledgments}

%%-------------------------------------------------
%%-------------------------------------------------
%\appendix

%\section{Symmetries of the $d$ function} \label{app}

%Here we list some of the symmetries of the $d$ function that are related to discussions in this paper,

% The \nocite command causes all entries in a bibliography to be printed out
% whether or not they are actually referenced in the text. This is appropriate
% for the sample file to show the different styles of references, but authors
% most likely will not want to use it.
%\nocite{*}

%\bibliography{reference} % Produces the bibliography via BibTeX.

%apsrev4-2.bst 2019-01-14 (MD) hand-edited version of apsrev4-1.bst
%Control: key (0)
%Control: author (8) initials jnrlst
%Control: editor formatted (1) identically to author
%Control: production of article title (0) allowed
%Control: page (0) single
%Control: year (1) truncated
%Control: production of eprint (0) enabled
%

\end{document}